# RF pulse amplifier for CVD-diamond particle detectors

C. Hoarau, G. Bosson, J.-L. Bouly, S. Curtoni, D. Dauvergne, P. Everaere, M.-L. Gallin-Martel, S. Marcatili, J.-F. Muraz, A. Portier and N. Rosuel.

Univ. Grenoble Alpes, Laboratoire de Physique Subatomique et Cosmologie – CNRS, 53 av. des Martyrs, 38000 Grenoble, France

*Abstract* – This article introduces a design of a Low Noise Amplifier (LNA), for the field of diamond particle detectors. This amplifier is described from simulation to measurements, which include pulses from α particles detection. In hadron therapy, with high-frequency pulsed particle beams, the diamond detector is a promising candidate for beam monitoring and time-stamping, with prerequisite of fast electronics. The LNA is designed with surface mounted components and RF layout techniques to control costs and to allow timing performance suitable for sub-nanosecond edges of pulses. Also this amplifier offers the possibility of high voltage biasing, a characteristic essential for driving diamond detectors. Finally the greatest asset of this study is certainly the minimization of the power consumption, which allows us to consider designs with multiple amplifiers, in limited space, for striped diamond detectors.

## I. Introduction

The treatment of tumors by carbon-ion or proton beams is an alternative cancer therapy known as hadron therapy. The ions deposit a large fraction of their dose at the end of their path (in the Bragg peak), in the patient's body. The advantage of this treatment is to have a higher dose deposition in the tumor volume and a lower one in the healthy nearby organs, compared to conventional radiotherapy. Since multiple sources of uncertainty on the ion range may cause deviations from the planned dose distribution [1], online control is desired. There are several hadron therapy monitoring techniques under development. One of them consists in detecting a fraction of prompt-gamma emitted in the energy range of 1 to 10 MeV by a fraction of the incident ions on their way through the patient body [2]. Time of Flight (ToF) measurements between the gamma detectors positioned around the patient and an interceptive beam monitor have demonstrated an improvement in the performance of Bragg peak localization. Such a beam tagging system is necessary to avoid continuous phase calibration of the HF signal of the accelerator. In this context, the LPSC team has initiated an innovative beam hodoscope project based on the use of diamond detectors [3].

In such an application, diamonds are used as solid state ionization chambers. Indeed, incident particles generate charge carriers (electron and hole pairs) on their path in the diamond material. Then the charge carriers drift under the influence of an external applied electrical field (typically 1 V/µm) and are collected by the electrodes located on top and bottom faces of the diamond material, inducing an electrical signal representative of the deposited energy. Two prototypes are underway. The first one will be made of a unique polycrystalline diamond sensor, $2 \times 2$ cm$^2$ × 300 µm, grown by means of the chemical vapor deposition method (pCVD). The second one will be made of a mosaic arrangement of four $0.45 \times 0.45$ cm$^2$ × 500 µm crystals sCVD (see figure 1).

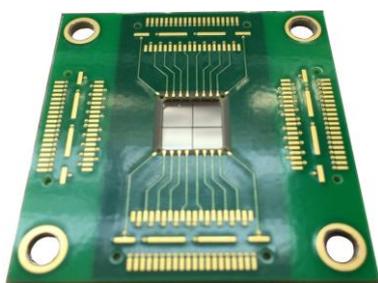

*Figure 1: Photography of diamond, 4 crystals of 4.5x4.5 mm$^2$, glued on its holder 5x5 cm$^2$, top view.*

Polycrystalline detectors have a lower charge collection efficiency than single crystal ones. However, the detection efficiency is good enough to detect proton bunches and individual carbon ions, whereas the timing resolution keeps at the scale of hundred picoseconds. In addition, only pCVD are presently available with large areas. The final version of the hodoscope, which will be available in the near future, will have a double-side strip metalization of each diamond sensor in order to localize the incident particle. The lift-off process is used to create the strips. Diamonds will be glued on a mezzanine board (see figure 1) and electrically connected by wire bonding. In the final version, the mezzanine board will be directly connected to the amplifiers board, each strip on diamond is read out by dedicated electronics. The number of strips is aimed to be at most 40, and the single channel voltage current preamplifier, presented in this paper, will be duplicated, for each strip, on one board, which justifies the development of discrete electronics in first intention to equip the beam monitors. In the present paper, we report on tests carried out solely with a unique sCVD diamond of 500 µm thickness from Element 6 [4], with 100 nm thick aluminum electrodes (see figure 2).

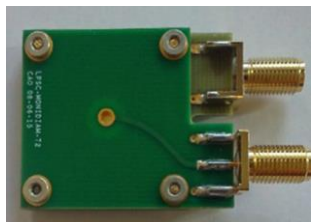

*Figure 2: Photography of diamond (at the center) stacked in its holder.*

Due to a high voltage bias, those charges are collected by strip electrodes on each side of the diamond. The amount of collected charge varies with the nature and energy of the impinging particle. It ranges from 7.5 fC for protons of 70 MeV detected by a 500 µm thick sCVD diamond, up to 520 fC for carbon ions of 95 MeV/nucleon detected in a 500 µm thick sCVD diamond. The signal is a pulse of a few nanoseconds, regards to the high mobility of charge carries in diamond. Hence, if we consider e.g. 60 fC charges generated, it induces a pulse of a few nanoseconds duration with a 12 µA amplitude at maximum on a 50 Ω load, and thus a 600 µV peak voltage. The signal is amplified by 40 dB. With a mobility of 2064 $cm^2.V^{-1}.s^{-1}$ for hole and 1714 $cm^2.V^{-1}.s^{-1}$ for electron in diamond [5], fast pulses are generated with a few 100 fs of rising time and few nanoseconds duration. To preserve the signal timing, a bandwidth over the gigahertz is required. In addition, to localize the generated pulse, each striped electrode should be amplified and so for an integration of multichannel amplifier, the power dissipation has to be limited to 75 mW per channel.

Therefore, we present the development of a fast electronics for amplifying pulses of charges from diamond detectors meeting those requirements. The prototype is a single channel preamplifier, which will be duplicated into the forty channels (on one board) for the final version of the beam hodoscope project. A cost effective design with surface-mount device, SMD, is detailed. Also frequency- and time-related measurements are presented.

## II. Design

Preamplifiers were designed with bipolar [6] [7], or JFET [8] transistors until 90's, when integrated circuit became popular. Integrated circuits were essentially CMOS devices [9], [10], but also bipolar [11], [12] and JFET and CMOS [13] combination. For low noise and low power application, discrete transistors came back around years 2000 with HEMT technology [14] [15] [16] [17] [18]. Nowadays there are still HEMT [19] existing, but a new technology emerges, the heterojunction bipolar transistor (HBT) with a slight improvement of the noise level, of the rising time and the radiation hardness [20]. Table 1 summarizes the performances of existing devices.

*Table 1 : Performance comparison. (a) Estimated, (b) from M. Fisher-Levine [19].*

| Ref. | Gain | Rise time | FWHM | ENC | SNR | Power |
|---|---|---|---|---|---|---|
| G. Bertuccio [14] | 31.5 dB (a) | 2.8 ns (b) | ~38 ns (a) | 139 $e^-$ | n/a | 0.7 W (a) |
| DBA II [16] | 42 dB | >0.5 ns | n/a | n/a | n/a | 1.44 W |
| CIVIDEC [17] | 60 dB | 350 ps | 1.38 ns | 1000 $e^-$ | MIP : 6.07:1 | 5.4 W (a) |
| DBA IV [21] | 50 dB | n/a | 10 ns | n/a | n/a | 1.8 W |
| M. Fisher-Levine [19] | 146 mV/fC | 2.12 ns | 3.75 ns | 293 $e^-$ | n/a | 1 W (a) |
| This work | 43 dB | 350 ps | 1.25 ns | 230 $e^-$ (a) | MIP : 4.3:1 | 75 mW |

The ENC is the equivalent noise charge at the input generating the output noise measured.

The amplifier design, presented in this article, is based on a first stage of a HBT in common emitter, to match timing and low noise level constraints, with a wideband NPN RF Si-Ge HBT BFP740. The second stage uses an all-integrated chip Low Noise Amplifier (LNA), the BGA427. Both devices are produced by Infineon. The BFP740 transistor has a 20 dB gain, with a transition frequency, $F_T$, about 39 GHz with a suitable biasing that stands a broadband amplification of 18 dB up to 3 GHz. The noise factor is under 0.6 dB up to 3 GHz for a collector current of 6 mA. At Infineon, there are two other candidates for the first stage, BFP840 and BFP842.

But there is no perfect transistor, it is a trade-off between gain, bandwidth, and noise factors. At another company, NXP Semiconductors, there is the BFU725F, which seems to be an equivalent transistor, but unfortunately the simulations tools are less supplied.

The other devices used in the design are SMD in 0603 package to minimize package parasitic effects such as parallel capacitor around 120 fF, and serial inductor about 140 pH.

The LNA schematic is presented in figure 3, compared to a classical common emitter amplifier, the design has a feedback resistor to limit the drifting of polarization point.

The detector is voltage-biased through a T shaped low-pass filter with serial resistors of 1 MΩ and a parallel capacitor of 1 nF. The detector signal is connected to the amplifier through a DC-block capacitor of 1 nF.

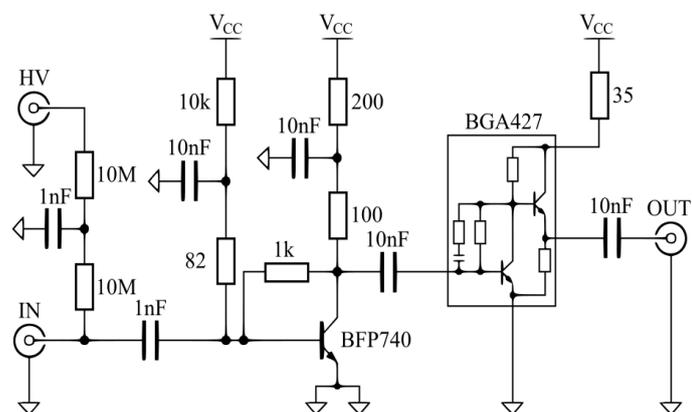

*Figure 3: Electronic schematic diagram of the LNA.*

Spice S-parameters simulations were conducted with LTSpice® from Analog devices. The standard Gummel and Poon model of transistors with package parasitic supplied by Infineon were used. In addition, the SMD components models included the parasitic effects of packaging described in the previous paragraph.

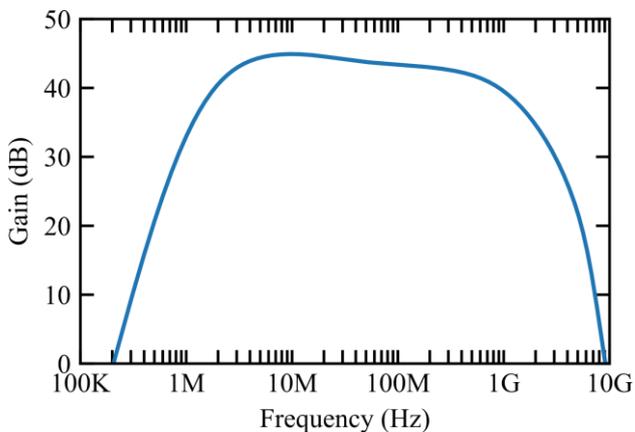

*Figure 4: Gain simulation (Spice).*

The gain curve, presented in figure 4, is not flat, it is 45 dB at 10 MHz and there is still 39.5 dB gain at 1 GHz, and 30.5 dB at 3 GHz.

Thanks to the Spice simulation, it was also possible to simulate and estimate the pulse response in time domain, so the graphic presented figure 5 shows the simulation response to an input pulse of 5 ns duration, 1 mV amplitude and with rising and falling time of 100 ps.

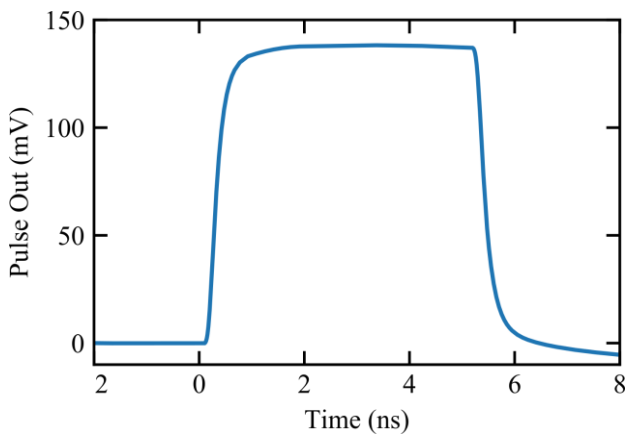

*Figure 5: Output pulse simulation (Spice).*

The rising time $T_{20/80}$ is 350 ps and the propagation delay is 750 ps, ± 10 ps. The gain is around 39 dB. Unfortunately, the spice model provided by Infineon, does not implement noise parameters, however the datasheet recommends to operate the transistor at $I_c$=6 mA to minimize the noise. The accumulation of noise figure through the two stages, is given by the Friis formula, with the noise figure of the first stage $F_1$, and the second stage $F_2$ and the gain of first stage $G_1$ :

$$F = F_1 + \frac{F_2 - 1}{G_1}$$

So the calculated noise figure is equal to 0.72 dB in the bandwidth, between 400 MHz and 4 GHz. Note that this characteristic gives the noise in excess, i.e. the degradation of the signal to noise ratio, but there is no information about the absolute noise in the datasheet, e.g. a noise quantified in nanovolt by square roots of Hertz.

The PCB layout shown in figure 6, is done for a FR4 substrate with 0.8 mm thickness. It has an effective permittivity of 3.9 and a dielectric tangent loss of 0.05. The coplanar grounded waveguide width is 1.2 mm and has a 1 mm gap to reach an characteristic impedance of about 55 Ω. This technological choice was made to reach a trade-off between width and characteristic impedance; the 50 Ω would have been obtained for a 1.5 mm width lines. A solid ground plane is at 0.8 mm.

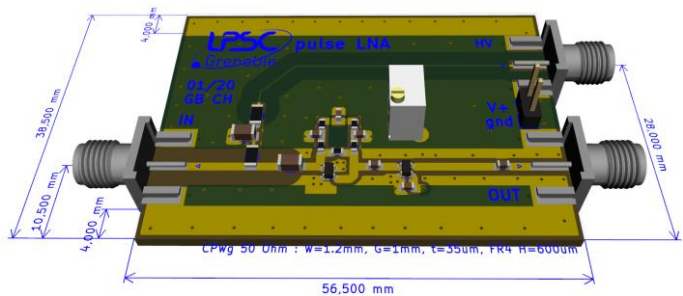

*Figure 6: Layout 3D drawing of LNA.*

The lines length are mimized and the active part surface is limited to 1.5 cm$^2$. On the other side (top layer) of the board voltage regulators trimmable with potentiometer are included to adjust biasing. The high voltage, i.e. ~500 V, line for the detector biasing is 3 mm spaced from ground to avoid sparking.

## III. Frequency domain measurements

For characterizing the amplifier, classical RF measurements are performed, starting with a vector network analyzer R&S ZNL6, to obtain the scattering parameters, i.e. gain $s_{21}$, and reflection coefficients $s_{11}$ and $s_{22}$, respectively at input and output. A compromised biasing for power consumption and gain is obtained with 3.8 V and 19 mA; so the total power consumption is 72 mW.

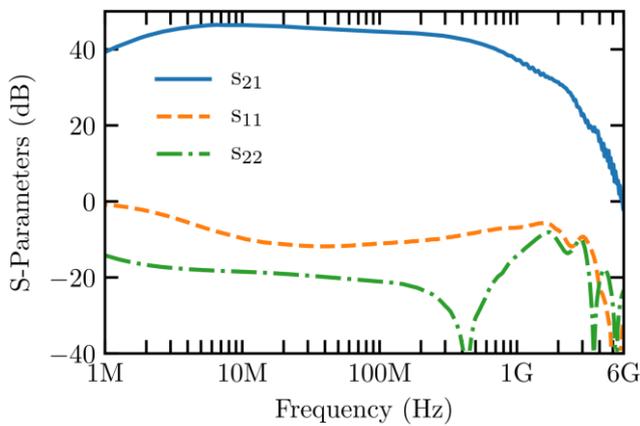
Figure 7: Scattering parameters of the LNA.

As predicted by the simulations shown in figure 7, the $s_{21}$ parameter is not flat through the frequency range, so the gain is 46 dB at 10 MHz, 31 dB at 2 GHz and still 23 dB at 3 GHz.

As the power consumption is proportional to the square of the voltage, for small signal approximation, it is possible to represent them on the same graph versus the gain (see figure 8).

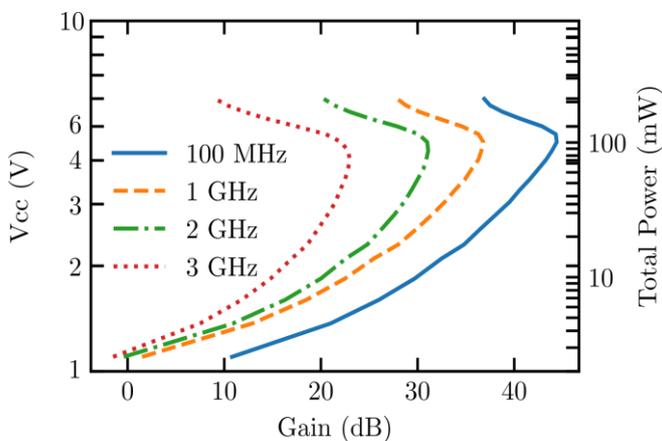
Figure 8: Gain function of the feeding voltage (left scale) or the power consumption (right scale).

The noise voltage density level is measured with the ZNL6 in spectrum mode with a 50 Ω at the input of the amplifier. The measurement reveal a noise level under $1\ nV/\sqrt{Hz}$ through a 2 GHz bandwidth (see figure 9).

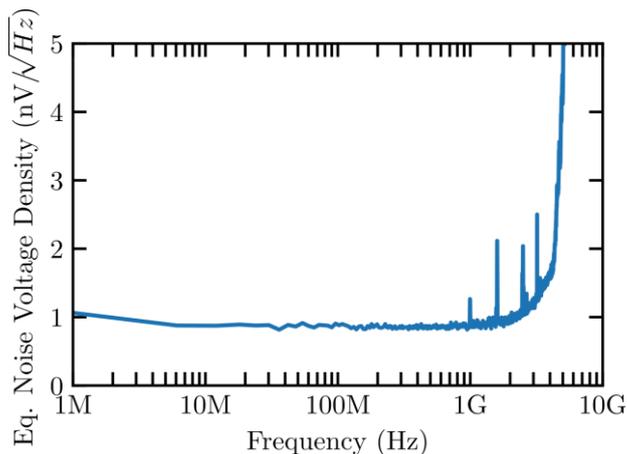
Figure 9: Equivalent input noise spectral density.

A measurement of the output root mean square noise voltage is done with a 50 Ω on input of the amplifier and a R&S 10K-2GHz URV3 millivoltmeter. The measure is 2.3 ± 0.05 mV. This measurement with following pulses measurements give an estimation of ENC of 230 e$^-$.

# IV. Time domain measurements

The time measurements presented were carried on two different benches, first, with a pulse generator, and second, with a $^{241}$Am α radioactive source and a diamond detector.

The pulse measurements, presented in figure 10, are performed with an Agilent pulse generator B1110A, with rising time of 0.8 ns, and a Tektronix scope 1 GHz at 5 Gsamples/s TDS5104.

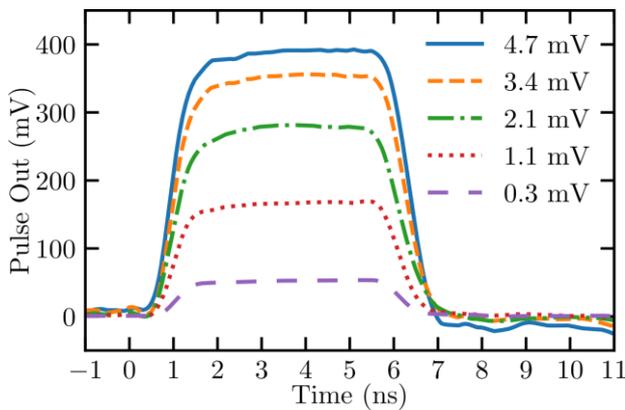

Figure 10: Oscillograms of output pulses functions of input amplitude ones with nominal values indicated in legend.

With a calibration measurement of the setup, and specifically each path delay, the propagation delay is deduced around to be 750 ps. With measurement of the gain function of the input signal's amplitude, it is possible to characterize the non-linearity with the 1 dB compression point which is around -50 dBm of input signal power. From the figure 10, it is possible to observe a baseline shift for the highest pulse amplitude due to the capacitive interconnections of the amplifiers stages.

For having equivalent condition measurement to our application, i.e. hadron therapy, we proceed to pulse measurement on a test bench equipped with a $^{241}$Am α source, and a diamond detector of 500 μm thickness (see figure 11).

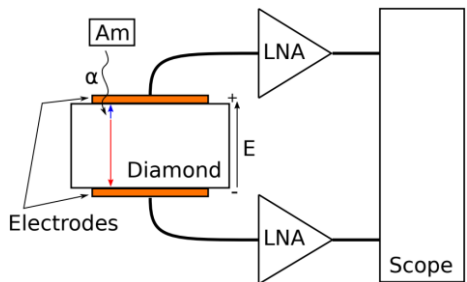

Figure 11: Setup measurement schematic.

The α particle generates charges in diamond CVD of 500 μm collected by high-voltage biased electrodes, i.e. 500 V, and amplified by our LNA to the oscilloscope, a LeCroy 4 GHz at 40 Gsamples/s, on one channel and by a CIVIDEC [22] one on the second channel. The disk metallized diamond is encapsulated by two printed circuits of FR4 board 50 Ω adapted with contact electrode connected to SMA connectors (figure 2).

The pulse measurements on each electrode are done with this setup, as illustrated in figure 12.

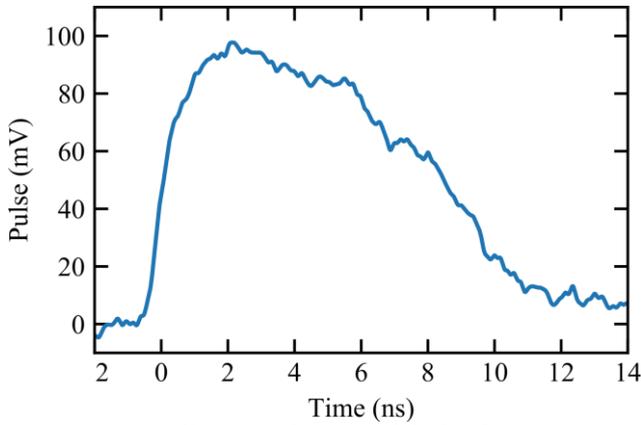

*Figure 12: Oscillogram of one amplified pulse.*

[241]Am nuclei emits α particles with 5.5 MeV energy. Each particle is totally absorbed within 12 μm of penetration. It generates 67 fC. The mean range of the α particles is then much smaller than the diamond thickness (300-500 μm). As a consequence, all the current due to the drift of charge carriers comes from a single species (either electrons or holes) traversing nearly the whole detector thickness. The shape of the pulse should be a perfect square according to the Ramo-Shockley theorem, but, as mentioned in [5], some imperfections in the diamond crystal change the charge carrier transport mechanism, and therefore change the shape of the pulse to a triangular. For instance, this is typical for pCVD detectors, for which the charge collection distance is of the order of ~100 μm. Note also that, in addition, if trapping occurs close to one surface, then the corresponding charge-up tends to decrease the bias electric field and decrease the drift velocity of the charge carrier. This effect is often referred to as polarization effect.

This measurement is repeated 1500 times for statistical extraction. The rising time 20-80% is measured at 760 ps. The jitter is evaluated as the difference of time between pulse from the reference amplifier (CIVIDEC) and the tested one. This difference is measured for few trigger-level fractions of the maximum. The jitter goes down to 43 ps for a trigger level of 25%. With a trigger of 50%, the jitter is 65 ps and at 75% it is 135 ps. Note also that the reference amplifier has a consumption of ~1 W, i.e. much higher than the design presented in this article.

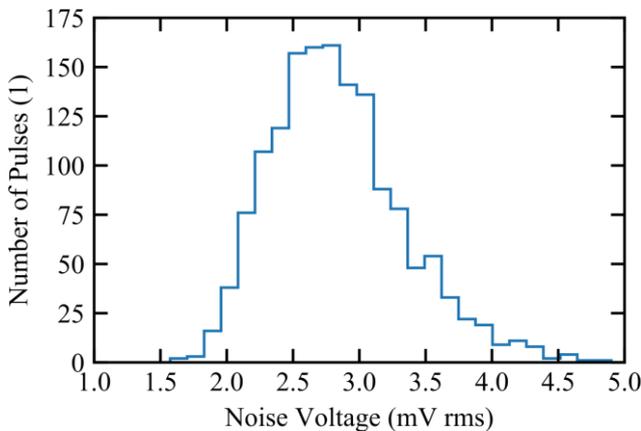

*Figure 13: Distribution of square root noise amplitude measurement for 1500 pulses.*

From oscillograms, it is possible to measure the noise level before and after the pulse and so to calculate the root mean square noise voltage. So the output total RMS noise voltage (detector and amplifier) is 2.84 mV with a σ of 0.43 mV (see figure 13).

## Outlook

This paper proposes a simple LNA architecture with RF methodology for amplifying pulses of particles detected with diamond CVD detectors. It has a 43 dB gain and a wideband of frequency with a gain of 31 dB at 2 GHz. The low power consumption, i.e. under 75 mW, makes it well suited for multi-channel readout of striped diamond detectors, at count rates greater than 100 MHz per channel and timing resolutions of the order of 135 ps to 43 ps for trigger level to 75% to 25% respectively. We are now developing a version with 40 channels on one board, with striped diamonds, a mosaic arrangement of four 0.45 × 0.45 cm² × 500 μm crystals sCVD, on a mezzanine board, to address the application of hadron therapy.